\documentclass{emulateapj}
\usepackage{amsmath,amssymb,graphicx,longtable}
\usepackage{natbib}
\usepackage[encapsulated]{CJK}
\usepackage[utf8x]{inputenc}

\newcommand{\hbeta}{H$\beta $ }
\newcommand{\halpha}{H$\alpha$ }
\newcommand{\itu}{{\it u }}

\shorttitle{{\it u}-band Emission as Star Formation Tracer}
\shortauthors{Zhou et al.}

\begin{document}
\begin{CJK}{UTF8}{gbsn}

	\title{SCUSS \itu Band Emission as a Star-Formation-Rate Indicator}
\author{Zhimin Zhou (周志民)\altaffilmark{1}, Xu Zhou\altaffilmark{1}, Hong Wu\altaffilmark{1}, Xiao-Hui Fan\altaffilmark{2}, Zhou Fan\altaffilmark{1}, Zhao-Ji Jiang\altaffilmark{1}, Yi-Peng Jing\altaffilmark{3}, Cheng Li\altaffilmark{4,5}, Michael Lesser\altaffilmark{2}, Lin-Hua Jiang\altaffilmark{6}, Jun Ma\altaffilmark{1}, Jun-Dan Nie\altaffilmark{1}, Shi-Yin Shen\altaffilmark{4}, Jia-Li Wang\altaffilmark{1}, Zhen-Yu Wu\altaffilmark{1}, Tian-Meng Zhang\altaffilmark{1}, and Hu Zou\altaffilmark{1}}

\altaffiltext{1}{Key Laboratory of Optical Astronomy, National Astronomical Observatories, Chinese Academy of Sciences, Beijing, 100012, China, zmzhou@bao.ac.cn}
\altaffiltext{2}{Steward Observatory, University of Arizona, Tucson, AZ 85721, USA}
\altaffiltext{3}{Center for Astronomy and Astrophysics, Department of Physics and Astronomy, Shanghai Jiao Tong University, Shanghai 200240, China}
\altaffiltext{4}{Shanghai Astronomical Observatory, Chinese Academy of Science, 80 Nandan Road, Shanghai 200030, P. R. China}
\altaffiltext{5}{Tsinghua Center for Astrophysics and Department of Physics, Tsinghua University, Beijing 100084, P. R. China}
\altaffiltext{6}{Kavli Institute for Astronomy and Astrophysics, Peking University, Beijing 100871, China}

\begin{abstract}
	We present and analyze the possibility of using optical {\it u}-band luminosities to estimate star-formation rates (SFRs) of galaxies based on the data from the South Galactic Cap {\it u } band Sky Survey (SCUSS), which provides a deep {\it u}-band photometric survey covering about 5000 $deg^2$ of the South Galactic Cap. Based on two samples of normal star-forming galaxies selected by the BPT diagram, we explore the correlations between {\it u}-band, H$\alpha$, and IR luminosities by combing SCUSS data with the Sloan Digital Sky Survey (SDSS) and {\it Wide-field Infrared Survey Explorer} ({\it WISE}). The attenuation-corrected {\it u}-band luminosities are tightly correlated with the Balmer decrement-corrected H$\alpha$ luminosities with an rms scatter of $\sim$ 0.17 dex. The IR-corrected \itu luminosities are derived based on the correlations between the attenuation of {\it u}-band luminosities and {\it WISE} 12 (or 22) $\mu$m luminosities, and then calibrated with the Balmer-corrected H$\alpha$ luminosities. The systematic residuals of these calibrations are tested against the physical properties over the ranges covered by our sample objects. We find that the best-fitting nonlinear relations are better than the linear ones and recommended to be applied in the measurement of SFRs. The systematic deviations mainly come from the pollution of old stellar population and the effect of dust extinction; therefore, a more detailed analysis is needed in the future work.

\end{abstract}

\section{Introduction}

Star formation is crucial in galaxy formation and evolution. Star-formation rate (SFR) is one of the most important diagnostics in understanding the evolution of galaxies across cosmic times. The measurement of SFRs in galaxies can give clues to the physical processes in shaping galaxies. As the improvements of new observations over the past decades, large-scale SFRs can be traced by many indicators at different wavelengths, ranging from the X-ray to the radio, and using both continuum and nebular recombination emission lines \citep[see][for reviews]{K98, Kennicutt2012}.

Among many SFR indicators, the ultraviolet (UV) continuum of galaxies is directly tied to the photospheric emission of young stellar population formed over the past 100 Myr, and hence serves as one of the most powerful probes of SFRs in galaxies \citep[e.g.,][]{K98, Hao2011, Murphy2011}. However, the main disadvantage of this tracer is its sensitivity to dust attenuation. Nebular recombination lines such as \halpha trace ionized gas surrounding stars with masses of $<$ 10$M_{\odot}$ and lifetimes of $\sim$ 3-10Myr, and represent a nearly instantaneous measure of the SFR \citep[e.g.,][]{Kewley2002, Brinchmann2004,Kennicutt2012}. The forbidden lines such as \mbox{[O {\sc ii}]} can be used as SFR tracers after empirical calibrations \citep[e.g.,][]{Kennicutt1992,Kewley2004}. However, the optical emission lines are sensitive to dust attenuation and the assumptions of the initial mass function (IMF). They are also affected by active galactic nuclei (AGNs) and metallicity \citep{Moustakas2006}. The starlight absorbed at UV and optical wavebands by the interstellar dust is re-emitted in infrared (IR), so the IR emission can be used as a sensitive tracer of the young stellar population formed over a period of 0$-$100 Myr \citep{Wu2005, Calzetti2007, Zhu2008, Wen2014, Ellison2016}. The major systematic errors of IR based SFRs come from the missing unattenuated starlight and dust heating from evolved stars. Besides that, composite multiwavelength tracers are applied to construct dust-corrected SFRs, such as the combination of UV and IR observations \citep[e.g.,][]{Fisher2009, Hao2011} and the combination of observed \halpha and IR luminosities \citep[e.g.,][]{Zhu2008, Kennicutt2009, Lee2013, Wen2014}.

Apart from the above tracers, the {\it u}-band ($\lambda \sim$ 355 nm) luminosity of a galaxy is dominated by young stars of ages $<$ 1 Gyr, and has been proved to be a reasonable star-formation indicator \citep{Cram1998, Hopkins2003, Moustakas2006, Prescott2009}.
Although the {\it u}-band traces more evolved stars than other tracers, its advantages over other tracers are that (1) high-quality images of {\it u}-band are easier to be obtained from ground-based telescopes, (2) large sky surveys can accumulate much more {\it u}-band photometric data than others in a short time. The early application of this method was analyzed with broadband luminosities of galaxies in the 1990s \citep[e.g.,][]{Cowie1997, Cram1998, K98}. Similar studies have been carried out with the image survey of the Sloan Digital Sky Survey \citep[SDSS;][]{York2000}. For example, \citet{Hopkins2003} constructed the calibration from {\it u}-band luminosity to SFR with a complete and volume-limited sample of star-forming galaxies from SDSS. With the SDSS sample, \citet{Moustakas2006} studied the effects of dust reddening and old stellar populations on the {\it U}-band luminosity as an SFR tracer, and presented the relation of {\it U} to H$\alpha$. \citet{Prescott2009} estimated the evolution of the cosmic star-formation history from {\it u}-band luminosity density after removing the contribution of old stellar populations and dust attenuation. 

The present calibrations of {\it u}-band-based SFRs are mainly constructed based on the SDSS data and in the form of single-band measurements. It is remarkable that the availability of recent and upcoming deep optical surveys, such as the Canada-France-Hawaii Telescope Legacy Survey, the South Galactic Cap {\it u}-band Sky Survey \citep[SCUSS;][]{ZhouX2016,Zou2015a}, and the Large Synoptic Survey Telescope \citep{Tyson2002}, provide opportunities to measure SFRs of unprecedented large samples with {\it u}-band data. With the advent of these deep and wide-field surveys, it becomes necessary to recalibrate the {\it u}-band-based SFR. 

This paper investigates the reliability and precision of {\it u}-band-based SFR using the observations from SCUSS. SCUSS provides deeper {\it u}-band photometric data than SDSS. These data are used to compare with \halpha and IR SFR indicators to derive an accurate {\it u}$-$SFR calibration for normal star-forming galaxies and to identify the limits of its applicability. We also derive calibrations of composite SFR indices, which combine {\it u}-band and IR luminosities to construct dust-corrected SFRs.

This paper is organized as follows. In Section \ref{sec:data} we briefly introduce the SCUSS survey and supplementary data and describe our samples. Section \ref{sec:analysis} compares the {\it u} band with \halpha and IR luminosities and derives their relations. In Section \ref{sec:discussion} we analyze the scatter and applicability limits of our SFR calibrations. In Section \ref{sec:summary} we summarize our results. Throughout this paper, we assume a flat $\Lambda$CDM cosmological model with $\Omega_M = $0.3, $\Omega_{\Lambda} = 0.7$, and $\it {H}_0 = 70 km s^{-1} Mpc^{-1}$.  

\section{Data and Sample}
\label{sec:data}
\subsection{SCUSS}
SCUSS is a {\it u}-band ($\lambda \sim$ 3538\AA) survey imaging $\sim$5,000 deg$^2$ in the northern part of the Southern Galactic Cap. It started in 2010 and finished in 2013. This survey was carried out using the 90Prime imager, a 1 deg$^2$ wide-field camera on the 2.3m Bok telescope at the Steward Observatory on Kitt Peak. The filter used is similar to that of the SDSS {\it u}-band, but a little bluer. The astrometric calibration is derived based on the Fourth US Naval Observatory CCD Astrograph Catalog (UCAC4), and the photometric calibration is carried out with SDSS Data Release 9 (DR9) photometric catalog. The limiting magnitude of SCUSS is $\sim$23.2 (with S/N=5 for point-like sources), about one magnitude deeper than SDSS {\it u}-band photometry. More detailed introduction of the SCUSS survey can be found in \citet{ZhouX2016} and the data reduction in \citet{Zou2015a}.

The survey area of SCUSS overlaps the photometric and spectroscopic survey of SDSS and the all-sky data of {\it Wide-field Infrared Survey Explorer} \citep[{\it WISE};][]{Wright2010} at 3.4, 4.6, 12, and 22 $\mu$m. These two surveys therefore provide spectroscopic and photometric data for our samples. 

\subsection{Supplementary data}
{\it SDSS } provides multi-color ({\it u, g, r, i, z}) images for one-third of the sky and spectra for millions of astronomical objects. The footprint of SDSS covers more than 75\% of SCUSS fields. In this paper, we use the SDSS DR9 photometric catalog and the emission-line measurements of galaxy spectra from MPA-JHU spectroscopic analysis \citep{Brinchmann2004, Tremonti2004}. The model magnitudes of galaxies are accepted (except {\it u}-band) in the photometric catalog, and the emission-line fluxes, which are corrected for Galactic extinction, are derived from the MPA-JHU catalogs.

The \halpha emission line is used as a reference SFR tracer to calibrate the {\it u}-band-based SFR indicators. To derive the \halpha luminosities and SFRs of galaxies, we need to correct the aperture bias and dust extinction for the \halpha emission line. The aperture correction is performed following the method of \citet{Hopkins2003} (their equation [A2]) by using the {\it r}-band fiber and Petrosian magnitudes of SDSS. The main assumption of this method is that the emission and star-formation activity through the fiber is characteristic of the whole galaxy, thus the curves-of-growth of the \halpha emission line and continuum across the galaxy are the same.
The line fluxes in the MPA-JHU catalog have been corrected for the foreground Galactic extinction by assuming the \citet{Cardelli1989} extinction curve ($R_V = 3.1$). To perform the intrinsic extinction correction, we use the \citet{Calzetti2000} extinction law ($R_V = 4.05$, $R_{H\alpha} = 3.32$ and $R_{H\beta} = 4.60$) and assume the Balmer decrement of intrinsic $H\alpha/$\hbeta $=$ 2.86 for the Case B recombination at $T_e = 10^4K$ and $n_e = 10^2 cm^{-3}$ \citep{Osterbrock1989}. Thus, the intrinsic extinction of \halpha is
\begin{equation}
	A_{H\alpha} = \frac{2.5R_{H\alpha}}{R_{H\beta} - R_{H\alpha}} log \bigg[\frac{(H{\alpha}/H{\beta})_{obs}}{2.86}\bigg].
\end{equation}

The SFRs of galaxies based on \halpha luminosities are derived using the calibration from \citet{K98}: $SFR_{H\alpha}(M_{\odot}\ yr^{-1})=7.9\times 10^{-42}[L(H\alpha)](erg\ s^{-1})$, which adopts a Salpeter IMF with stellar masses in the range of 0.1--100 M$_{\odot}$ and stellar population models with solar abundances.

\begin{figure*}
	\centering
	\includegraphics[width=0.4\hsize]{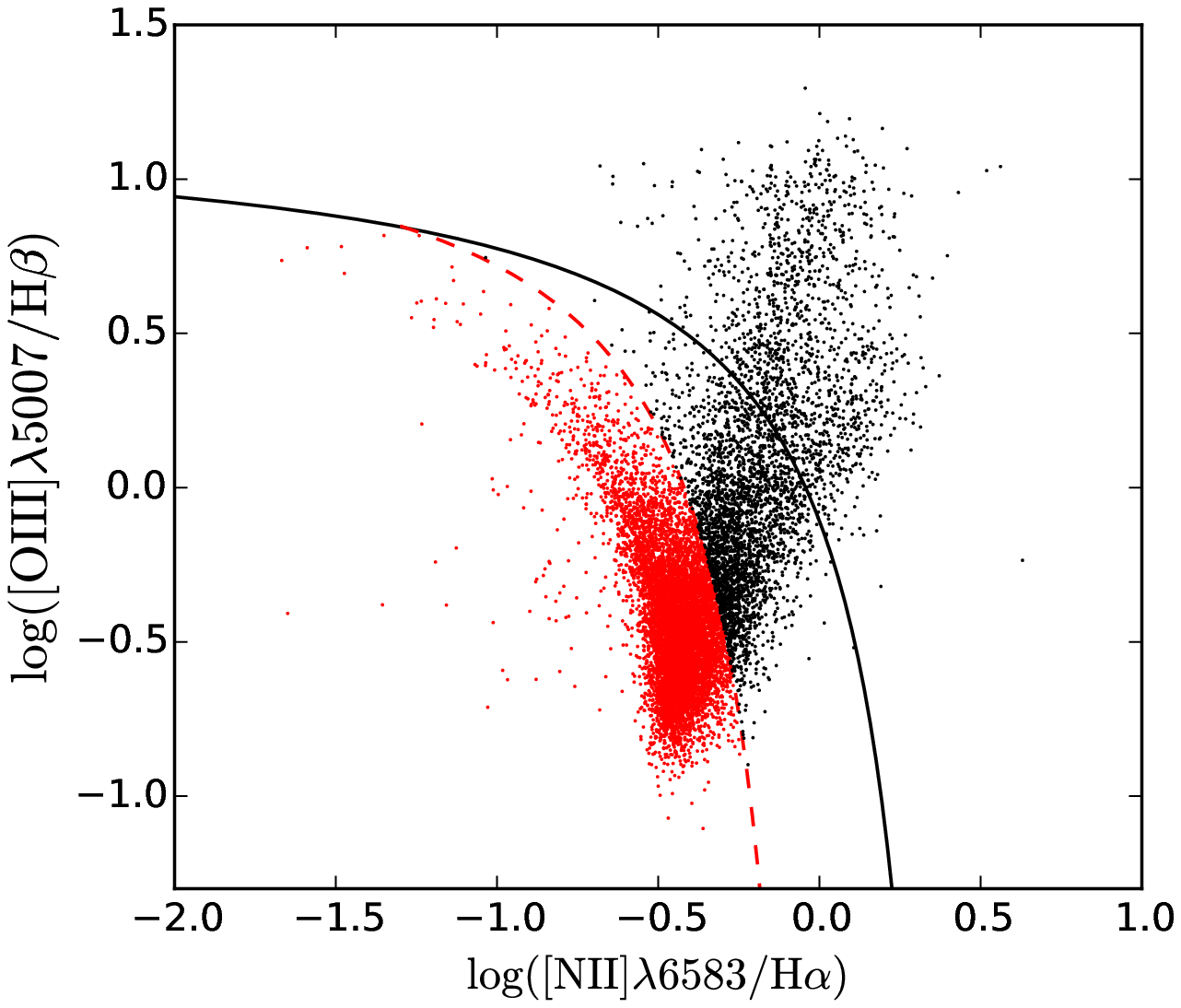}
	\includegraphics[width=0.4\hsize]{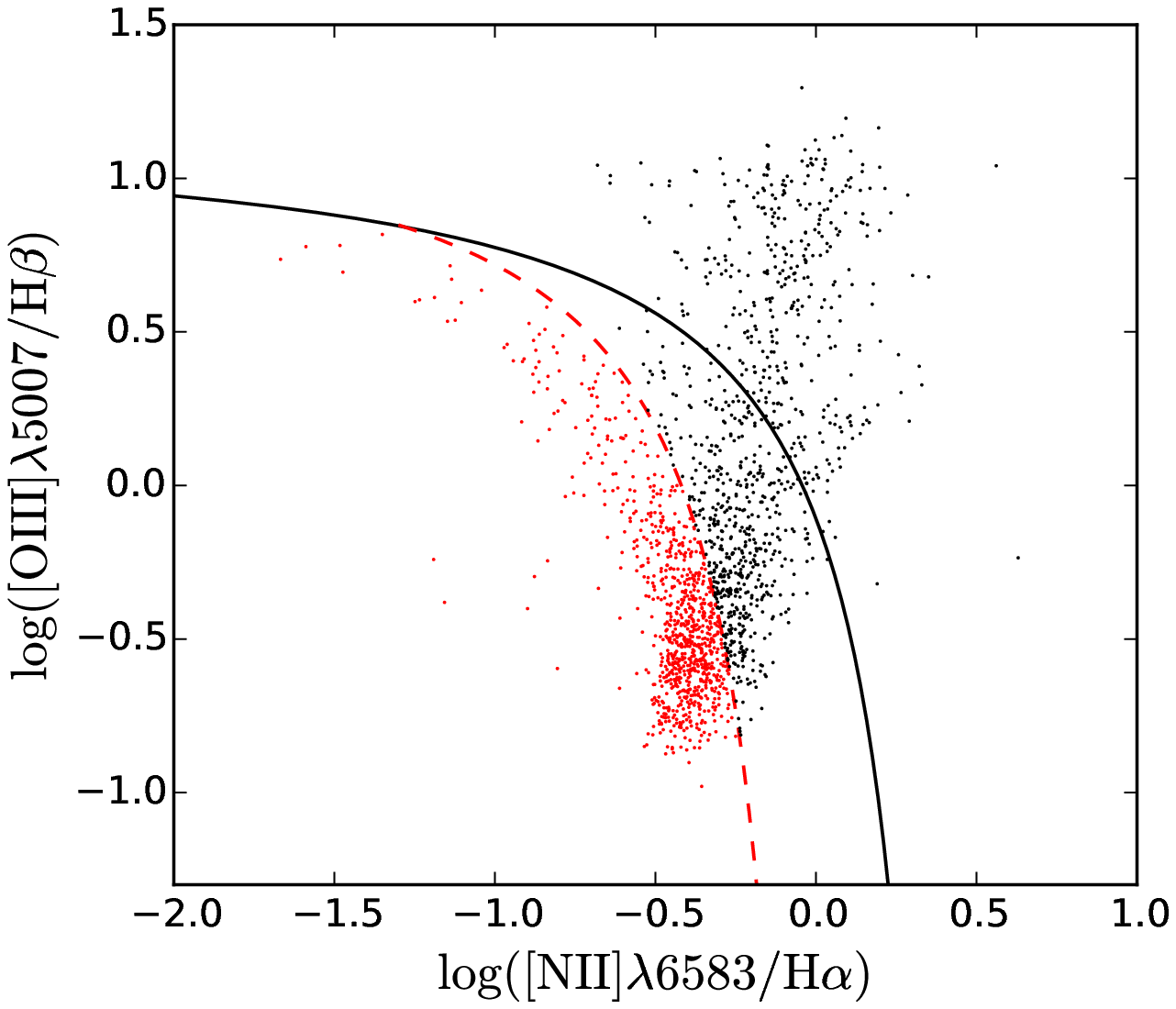}
	\caption{BPT diagnostic diagrams used for selecting star-forming samples. The sources in the panels are selected by cross-matching of SCUSS, SDSS MPA-JHU and ALLWISE catalogs, and have all diagnostic emission lines detected at S/N $>$ 3 and S/N $\ge$ 10 in the 12 $\mu$m band ({\it left}), S/N $\ge$ 10 in the 22 $\mu$m band ({\it right}). The demarcations between star-forming galaxies and AGNs are from \citet{Kauffmann2003} (red dashed curve) and \citet{Kewley2001} (black solid curve). The objects (red points) below the red dashed curve are the {\it pure} star-forming galaxies with no contamination of AGNs, and are selected as the Sample 1 ({\it left}) and Sample 2 ({\it right}). There are 8724 galaxies in Sample 1, and 872 in Sample 2.}
	\label{fig_BPT}
\end{figure*}

{\it The Wide Field Infrared Survey Explorer} \citep[{\it WISE};][]{Wright2010} provides a mid-infrared imaging survey of the entire sky at 3.4, 4.6, 12, and 22 $\mu$m with an angular resolution of 6\farcs1, 6\farcs4, 6\farcs5, and 12\farcs0, respectively. WISE 12 and 22$\mu$m luminosities have proved to be useful SFR indicators \citep{Jarrett2013, Lee2013, Wen2014}, and WISE 3.4$\mu$m luminosity can be used to calculate the stellar masses of galaxies \citep{Wen2013}. We use the ALLWISE source catalog which contains positional and photometric information for over 747 million objects detected on the deep, coadded WISE Images. The profile-fit photometry magnitude of each band is accepted in our research.

\begin{figure*}
	\centering
	\includegraphics[width=\hsize]{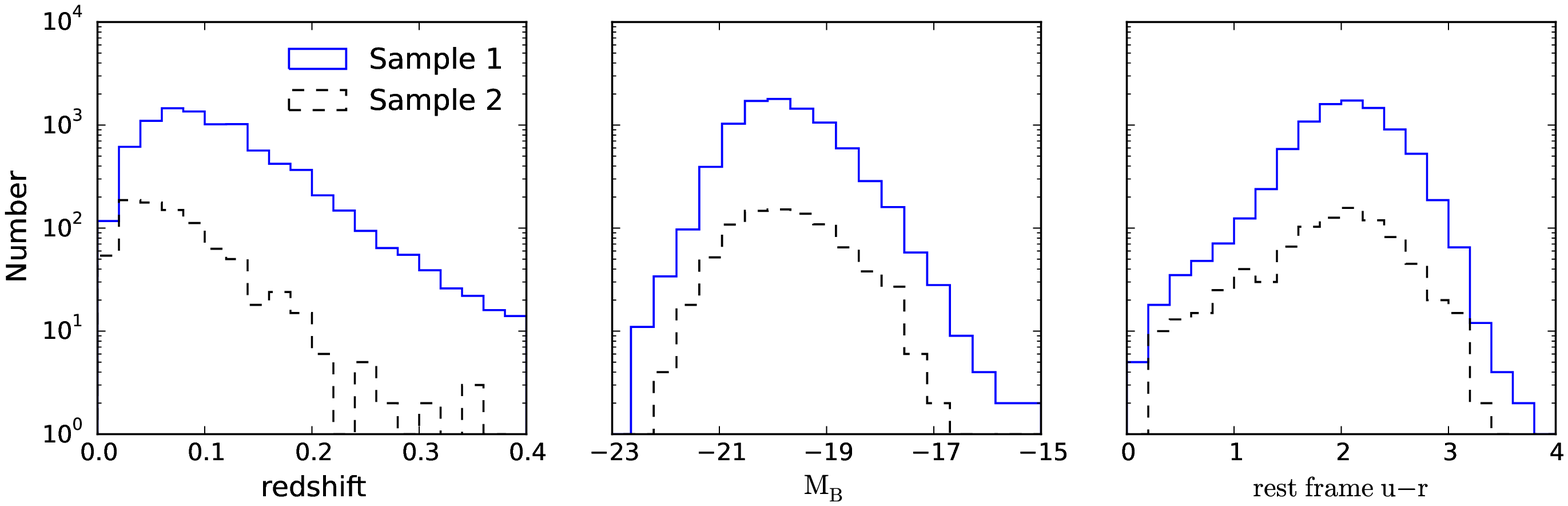}
	\caption{Distribution of redshifts, B-band absolute magnitudes ($M_B$) and rest-frame u$-$r colors for Sample 1 (blue solid line) and Sample 2 (black dashed line). The redshift is derived from the MPA-JHU catalogs and $M_B$ is calculated from the {\it K}-correction SDSS {\it g}- and {\it r}-band magnitudes based on the transformation equation of \citet{Smith2002}.}
	\label{hist1}
\end{figure*}
\begin{figure*}
	\centering
	\includegraphics[width=\hsize]{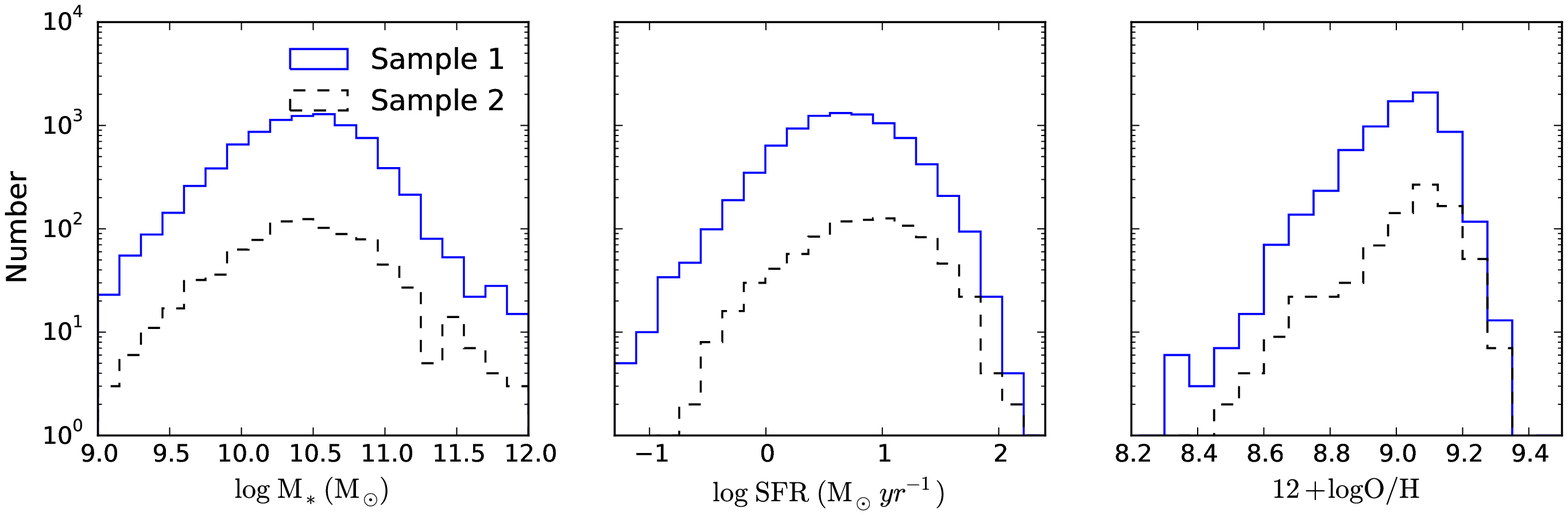}
	\caption{Distribution of stellar masses, SFRs and gas-phase metallicities for Sample 1 (blue solid line) and Sample 2 (black dashed line). The stellar masses of galaxies are estimated with the WISE 3.4 $\mu$m luminosities, the SFRs are calculated with the aperture- and extinction-corrected \halpha luminosities, and the metallicities are derived from the SDSS MPA-JHU catalog.}
	\label{hist2}
\end{figure*}

\subsection{Sample}
\label{sec:sample}

The SCUSS database has been cross-matched and integrated with the SDSS DR9 photometric catalog with a matching radius of 2\farcs0 \citep{Zou2015b}. We match the SCUSS sources with the MPA-JHU catalog to get the spectroscopic information, and then match with the ALLWISE catalog with 3\farcs0 radius. If multiple optical sources are located within 3\farcs0 of a WISE source, the nearest optical source is selected. In order to produce the accurate SFR calibration, we select the {\it pure} star-forming galaxies with no contamination of AGN based on the BPT diagnostic diagram: \mbox{[N {\sc ii}]}/\halpha versus \mbox{[O {\sc iii}]}/\hbeta \citep{Baldwin1981, Veilleux1987}. In Figure \ref{fig_BPT}, we select the sources that have all BPT diagnostic lines detected at S/N $>$ 3 and lie below the AGN rejection line of \citet{Kauffmann2003}:
\begin{equation}
	\log([O III]/H\beta) \leq {0.61 \over \log([N II]/H\alpha)-0.05} + 1.3.
\end{equation}
From the star-forming galaxies, we select galaxies with S/N $\ge$ 10 in the 12 $\mu$m band as Sample 1, and select galaxies with S/N $\ge$ 10 in the 22 $\mu$m band as Sample 2. There are 8724 galaxies in Sample 1 and 872 in Sample 2.

To derive the rest-frame magnitudes and color, we compute the {\it K}-correction for each galaxy using the spectral energy distribution (SED) templates and fitting code of \citet{Assef2010}. The SED templates are built for AGNs and galaxies in the wavelength range from 0.03 to 30 $\mu$m. We use the {\it model} magnitudes of SCUSS {\it u}-band and SDSS {\it griz} bands and the profile-fit photometry magnitudes of WISE W1--4 bands for the SED fit during the {\it K}-correction.

Figure \ref{hist1} shows the distributions of redshifts, B-band absolute magnitudes ($M_B$) and rest-frame $u-r$ colors for the two samples. The redshift information is derived from the MPA-JHU catalogs and $M_B$ is calculated from the SDSS {\it g}- and {\it r}-band magnitudes after K-correction based on the transformation equation of \citet{Smith2002}. We find no significant difference in the distributions of the two samples. Most of the galaxies have redshifts less than 0.4 and cover a $M_B$ range of $-23 < M_B < -16$.
 
Figure \ref{hist2} shows the distribution of stellar masses, SFRs, and gas-phase metallicities. The stellar masses of galaxies are estimated with WISE 3.4 $\mu$m luminosities by using the calibration of \citet{Wen2013} (their coefficients for H {\sc ii} galaxies), and the metallicities are derived from SDSS MPA-JHU catalog. No obvious difference is found in these distributions for the two samples, which have the stellar masses of $10^9$ -- $10^{12} M_{\odot}$, SFRs of 0.1 -- 100 $M_{\odot} yr^{-1}$, and 12+log O/H of 8.3 -- 9.3.

\begin{deluxetable*}{ccccccc}
 \tablewidth{0pt}
 \tablecaption{Correlation Coefficients between \itu and \halpha and IR
 \label{tab1}}
 \tablehead{\colhead{y} & \colhead{x} &	\colhead{a} & \colhead{b} & \colhead{s} & \colhead{c} & \colhead{r} \\
 \colhead{(1)} & \colhead{(2)} & \colhead{(3)} & \colhead{(4)} & \colhead{(5)} & \colhead{(6)} & \colhead{(7)} }
 \startdata
 $L({\it u})_{corr}$ & $L(H\alpha)_{corr}$ & 10.31$\pm$0.16 & 0.80$\pm$0.01 & 0.17 & 1.81$\pm$0.13 & 0.88\\
 $L({\it u})_{corr}$ & L(12 $\mu$m) & 11.79$\pm$0.17 & 0.74$\pm$0.01 & 0.18 & 0.50$\pm$0.13 & 0.87\\
 $L({\it u})_{corr}$ & L(12 $\mu$m) & 19.35$\pm$0.68 & 0.56$\pm$0.02 & 0.27 & 0.35$\pm$0.29 & 0.73\\
 $L({\it u})_{obs}+1.83\times L(12 \mu m)$ & $L(H\alpha)_{corr}$ & 9.78$\pm$0.16 & 0.81$\pm$0.01 & 0.16 & 1.81$\pm$0.13 & 0.89\\
 $L({\it u})_{obs}+1.46\times L(22 \mu m)$ & $L(H\alpha)_{corr}$ & 9.26$\pm$0.59 & 0.82$\pm$0.01 & 0.19 & 1.68$\pm$0.16 & 0.83
 \enddata
 \tablecomments{Cols (1) and (2): the names of wavelength-band and emission-line luminosities, which are in units of erg $s^{-1}$. (3) and (4): The coefficients a and b of the nonlinear fit: log(y) = a + b $\times$ log(x). (5): The standard deviation s of the fitting residuals. (6): The coefficient c of the linear fit: log(y) = c + log(x). (7): The Spearman rank-order correlation coefficient r.}
\end{deluxetable*} 

\begin{figure}
	\centering
	\includegraphics[width=\columnwidth]{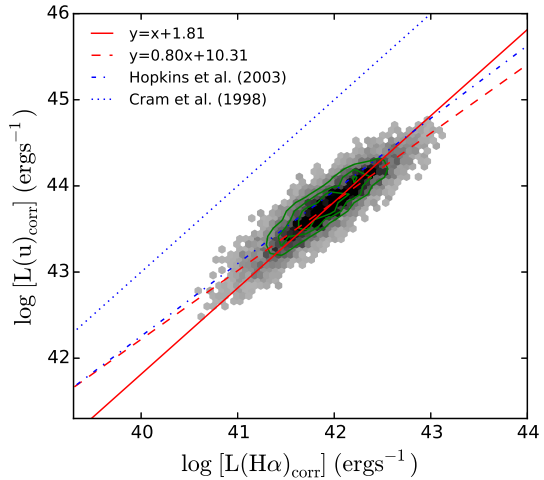}
	\caption{Comparison of the luminosities of {\it u}-band and \halpha for Sample 1. The \halpha luminosities are derived from the MPA-JHU catalog and are corrected for the dust extinction and aperture bias. The {\it u}-band luminosities are also corrected for the dust extinction. The green contours are number densities of the sample distribution. The red solid and dashed lines represent the best linear and nonlinear fits in this work. The blue dashed-dotted line represents the relation by \citet{Hopkins2003} using SDSS DR1. The dotted line is the relation derived by \citet{Cram1998} using the {\it U}-band data derived from relevant synthesized spectra.}
	\label{u_vs_Ha}
\end{figure}

\section{Analysis}
\label{sec:analysis}

\subsection{Dust correction to {\it u}-band Luminosities}

As addressed in the Introduction, {\it u}-band luminosity can be used as an SFR indicator of galaxies, but similar to UV, it also suffers from dust extinction. Therefore, it is necessary to correct the extinction for the {\it u}-band flux. The Galactic extinction in the SCUSS {\it u}-band follows the value in SDSS {\it extinction\_u}, which is derived from the \citet{Schlegel1998} dust maps and the assumption of $R_V = 3.1$ absorbing medium. 

The intrinsic dust extinction for {\it u}-band luminosities is estimated using the \citet{Calzetti2000} obscuration curve and the color excess of stellar continuum $E_s(B-V)$. Since the geometries of the dust distribution are different relative to stars and gas, and the covering factor of the dust is smaller for stars than for gas, the stellar continuum can be subject to a lower degree of reddening than nebular emission lines \citep{Calzetti1997, Johansson2016}, and $E_s(B-V)$ should be smaller than the color excess of ionized gas $E_g(B-V)$. Here we use the empirical relation of \citet{Calzetti2000}: $E_s(B-V) = 0.44E_g(B-V)$, where $E_g(B-V)$ is derived based on the Balmer decrements. This yields the relation of $A_{\it u}=0.81A_{H\alpha}$. The attenuation-corrected {\it u}-band luminosities $L({\it u})_{corr}$ is
\begin{equation}
	L({\it u})_{corr}=L({\it u})_{obs} \times 10^{0.4A_{\it u}},
	\label{eq0}
\end{equation}
and the amount of attenuation at the {\it u}-band is
\begin{equation}
	L({\it u})_{att}=L({\it u})_{corr} \times (1-10^{-0.4A_{\it u}}).
	\label{eq1}
\end{equation}

\subsection{{\it u} band versus \halpha}
\label{subsec:u_ha}

Because nearly all galaxies in Sample 2 are included in Sample 1, we use Sample 1 to compare the luminosities of the \itu band and the \halpha emission line, as shown in Figure \ref{u_vs_Ha}. It can be clearly seen that the attenuation-corrected {\it u}-band luminosities correlate tightly with the Balmer decrement-corrected \halpha luminosities with the Spearman rank-order correlation coefficient of 0.88. The best linear and nonlinear fits to Sample 1 are shown as:
\begin{equation}
	log~L({\it u})_{corr}= log~L(H\alpha)_{corr} + (1.81\pm0.13), \label{eq_liner_u_ha}
\end{equation}

\begin{equation}
	log~L({\it u})_{corr}= (0.80\pm0.01) \times log~L(H\alpha)_{corr} + (10.31\pm0.16).
	\label{eq_noliner_u_ha}
\end{equation}
The scatter of the nonlinear fit is $\sim$ 0.17 dex. For comparison, the relations from \citet{Hopkins2003} and \citet{Cram1998} are also plotted in Figure \ref{u_vs_Ha}. Our fitting results are similar to that of \citet{Hopkins2003}, which used SDSS data and the same dust correction methods. However, the relation of \citet{Cram1998} deviates from our results; the reason is likely that the data of \citet{Cram1998} are the broadband {\it U} (not {\it u}) luminosities of disk galaxies scaled from far-UV using relevant synthesized spectra.

\begin{figure*}
	\centering
	\includegraphics[width=0.4\hsize]{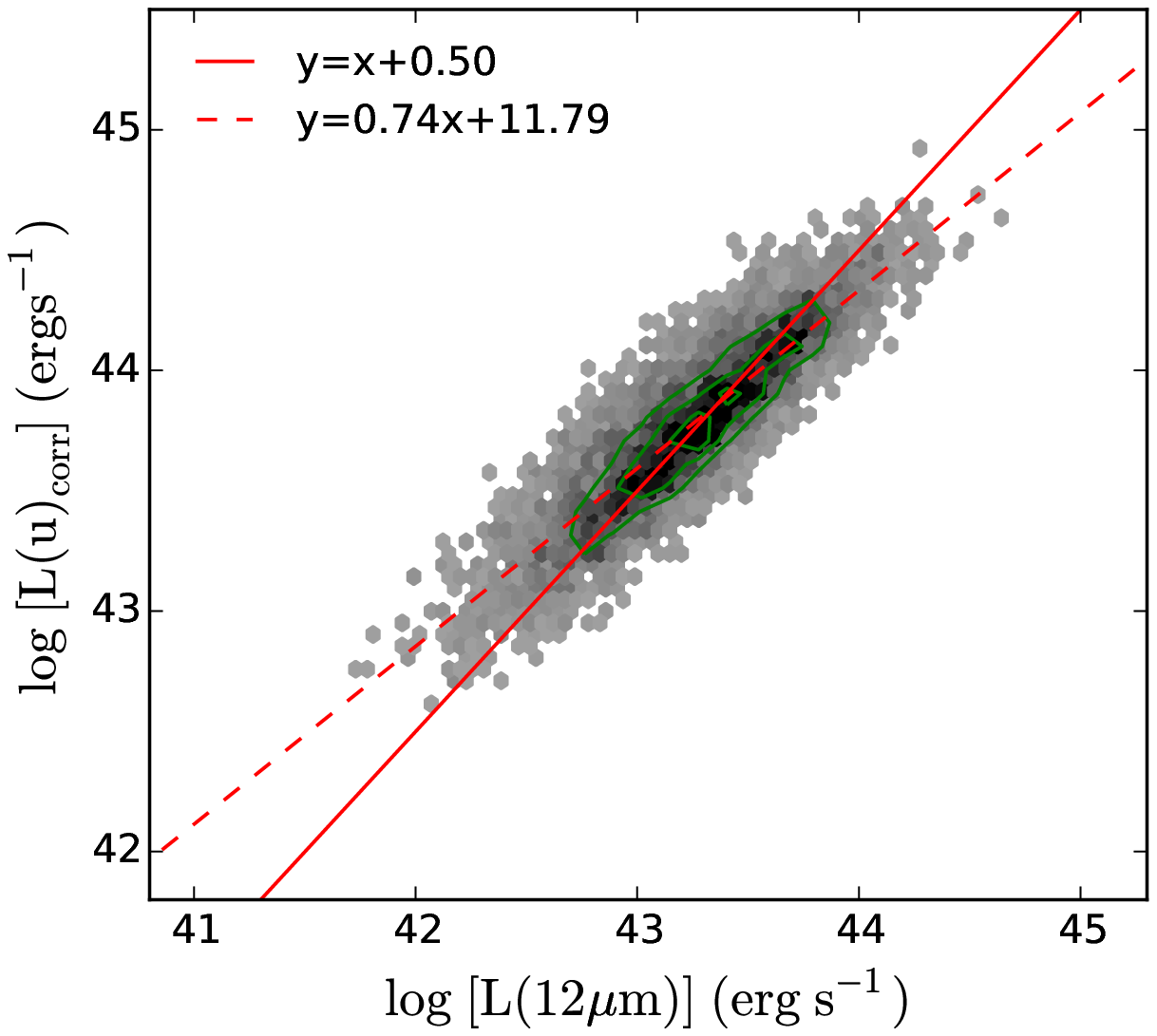}
	\includegraphics[width=0.4\hsize]{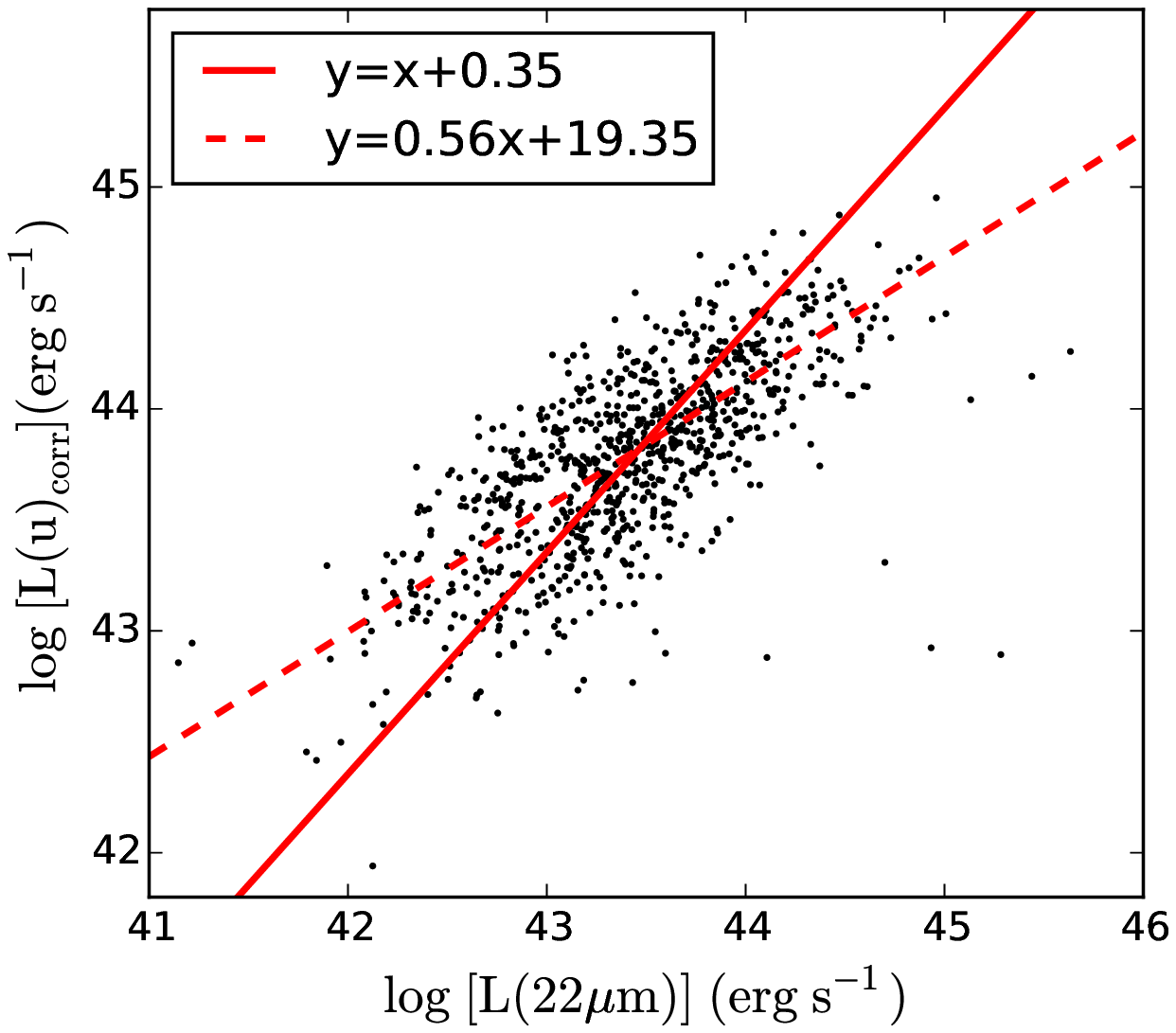}
	\caption{{\it u}-band luminosities as a function of WISE mid-IR luminosities. The {\it u}-band luminosities are corrected for dust extinction. {\it Left}: {\it u}-band versus WISE 12 $\mu$m. The objects in Sample 1 are used with the green counters for the number densities in the plot. {\it right}: {\it u}-band versus WISE 22 $\mu$m. The objects in Sample 2 are used. In each panel, the best linear and nonlinear fits are shown with solid and dashed lines, respectively.}
	\label{u_vs_wise}
\end{figure*}

\begin{figure*}
	\centering
	\includegraphics[width=0.4\hsize]{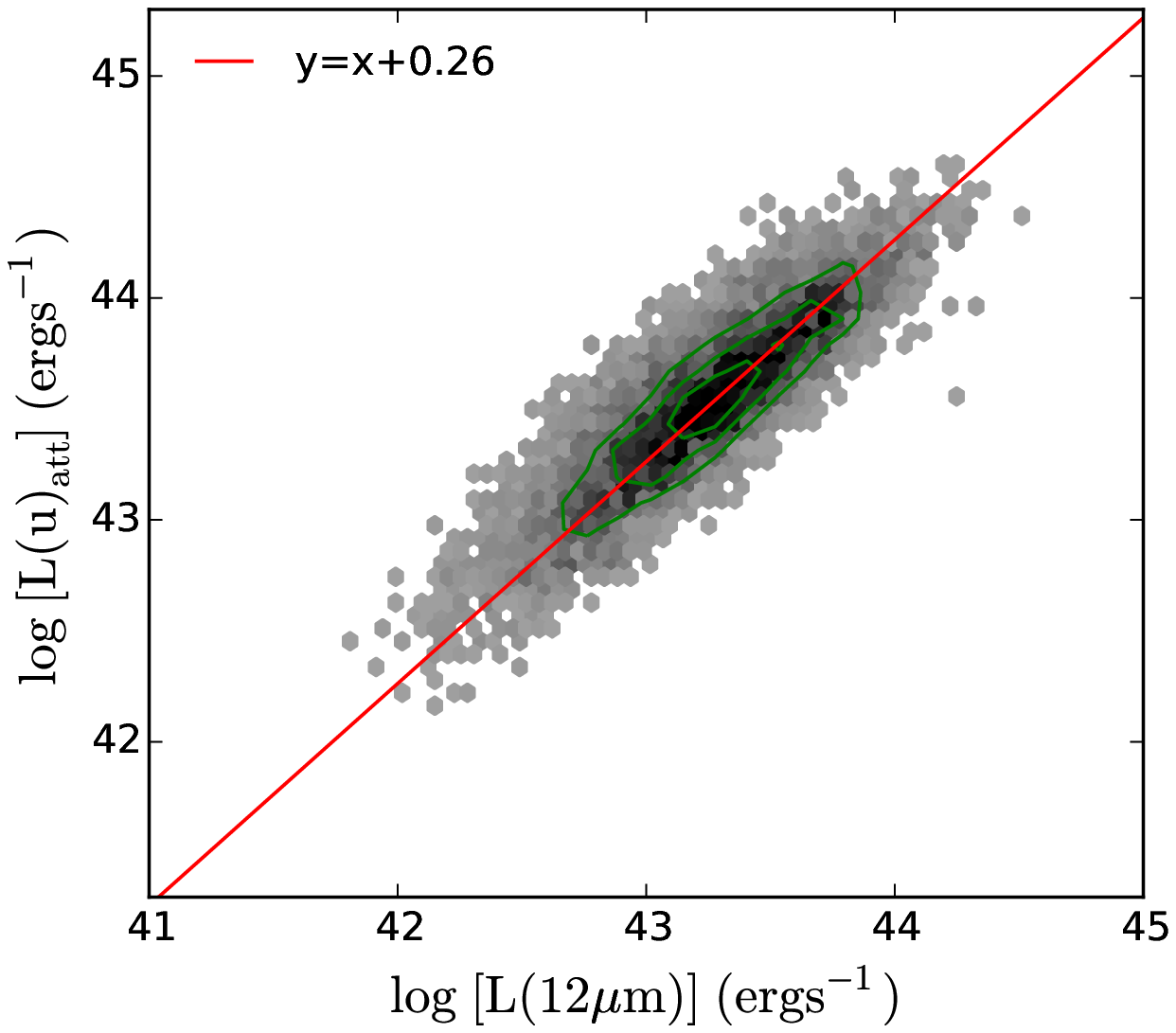}
	\includegraphics[width=0.4\hsize]{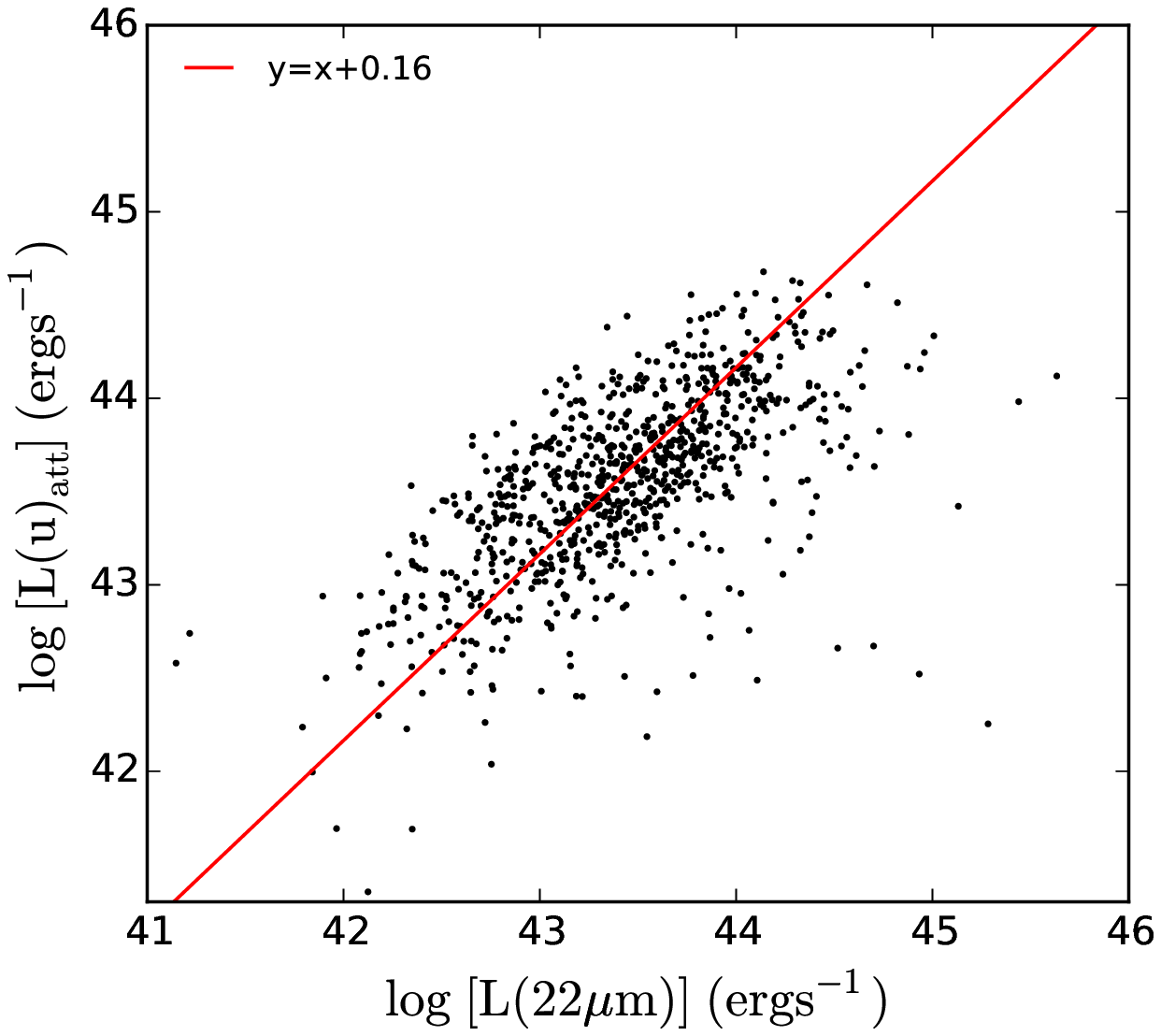}
	\caption{Correlating the attenuation of {\it u}-band luminosities with WISE mid-IR luminosities. The attenuation of {\it u}-band luminosities is the starlight in {\it u}-band absorbed by interstellar dust and is calculated using Eq.~\ref{eq1}. The {\it left} panel is for WISE 12 $\mu$m using Sample 1 along with the green counters for the number densities of the sample distribution, and the {\it right} panel is for WISE 22 $\mu$m using Sample 2. The best linear fits are shown with solid lines in both panels.}
	\label{uext_vs_wise}
\end{figure*}

\begin{figure*}
	\centering
	\includegraphics[width=0.4\hsize]{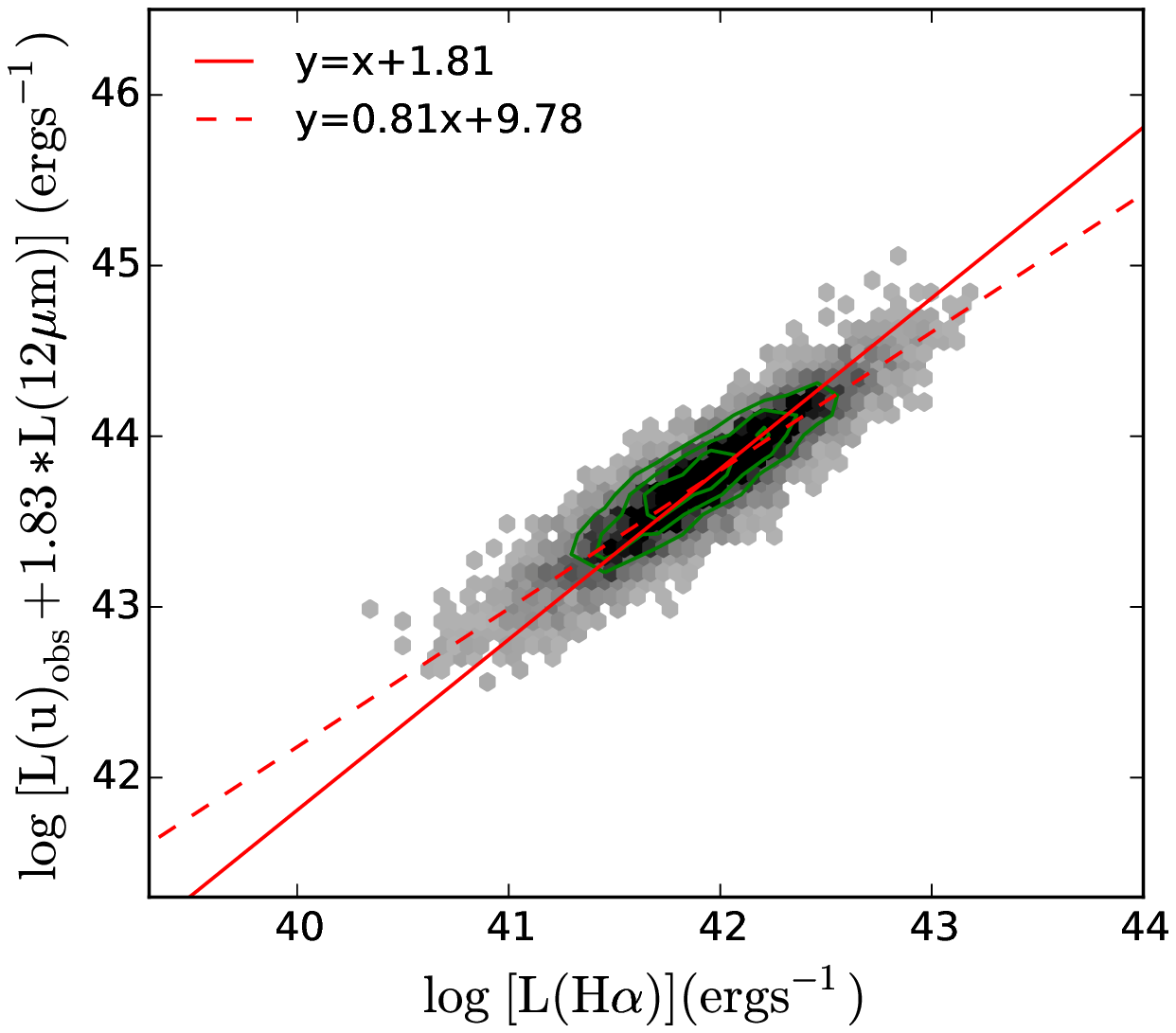}
	\includegraphics[width=0.4\hsize]{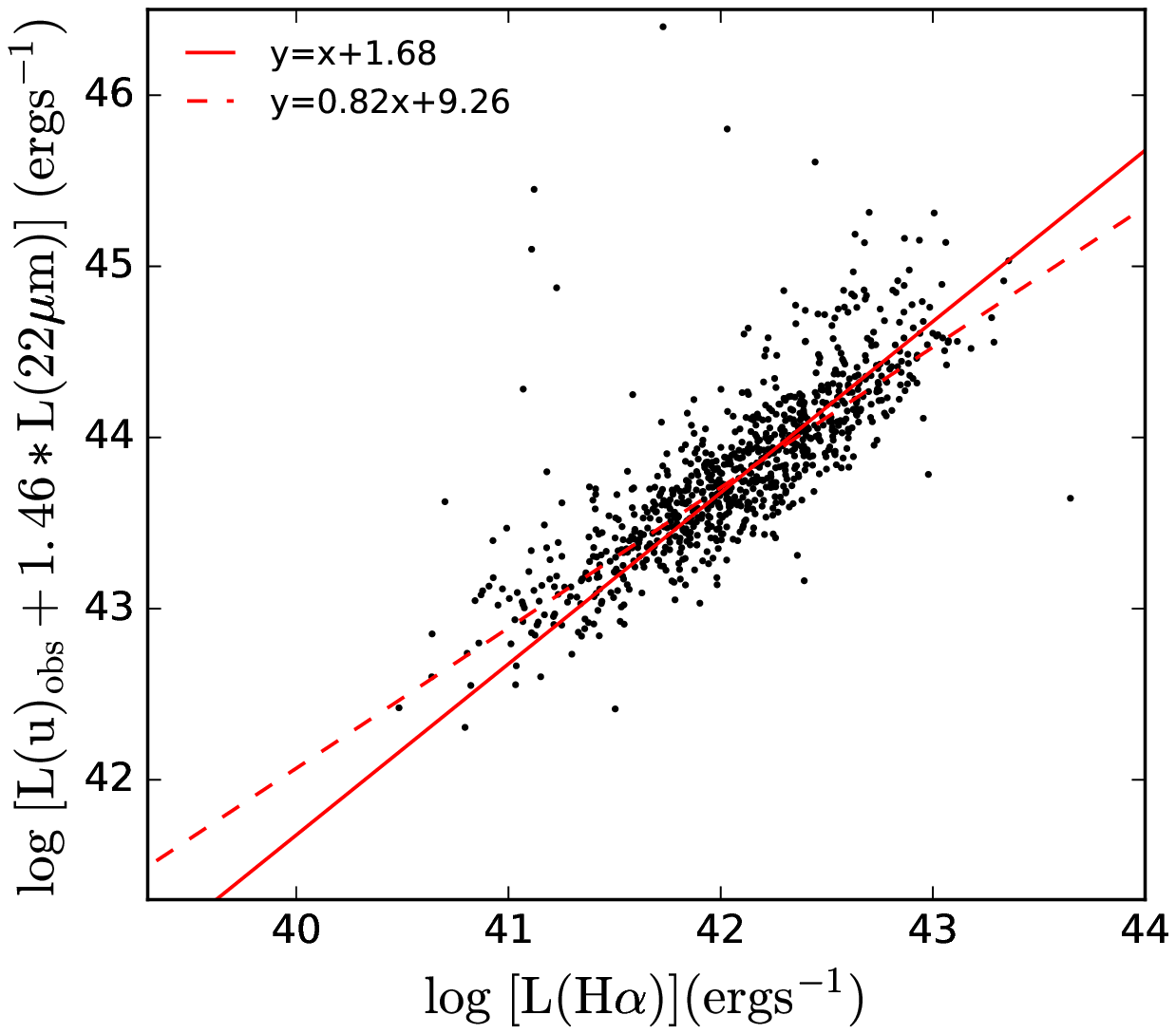}
	\caption{IR-corrected {\it u}-band luminosities as a function of \halpha luminosities. We use the attenuation of the \itu band versus WISE 12/22 $\mu$m relation in Figure~\ref{uext_vs_wise} to get the IR-corrected {\it u}-band luminosities (Eq.~\ref{IR12u} and \ref{IR22u}). The \halpha luminosities are Balmer decrement-corrected and aperture-bias corrected. The {\it left} panel shows the \itu$+$12 $\mu$m versus \halpha relation for Sample 1, the green counters are number densities of the sample distribution. The {\it right} panel shows the \itu$+$22 $\mu$m versus \halpha relation for Sample 2. In both panels, the best linear and nonlinear fits are shown with solid and dashed lines, respectively.}
	\label{uIR_vs_Ha}
\end{figure*}

\subsection{{\it u} band versus WISE 12 and 22 $\mu$m}

\begin{figure*}
	\centering
	\includegraphics[width=0.9\hsize]{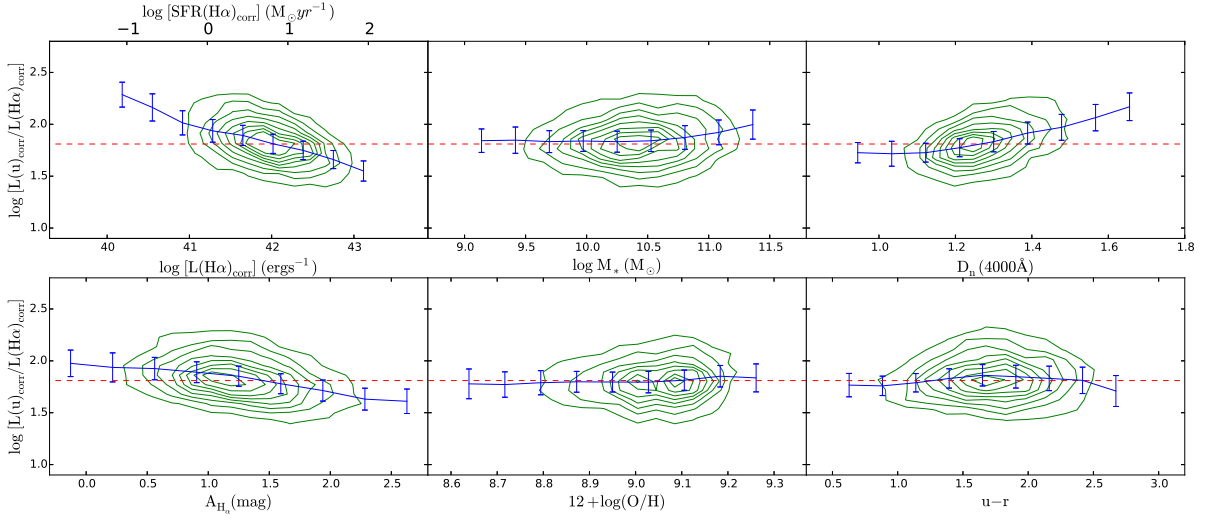}
	\caption{Logarithm residuals of {\it u}-band luminosities $L({\it u})_{corr}$ relative to H$\alpha$ luminosities $L(H\alpha)_{corr}$ as functions of Balmer-attenuation-corrected H$\alpha$ luminosities, stellar masses of galaxies, 4000\AA~ break $D_n(4000\AA)$, dust attenuation in H$\alpha$, gas-phase oxygen abundance, rest-frame u$-$r color from left to right and top to bottom, respectively. $L({\it u})_{corr}$ and $L(H\alpha)_{corr}$ are both attenuation-corrected luminosities. The stellar masses are measured based on WISE 3.4 $\mu$m luminosities. The $D_n(4000\AA)$ and gas-phase oxygen abundance are derived from MPA-JHU catalogs. The green contours are number densities of the sample distribution. The blue error bar represents the median value and standard deviation in each bin.}
	\label{comp1_u_Ha}
\end{figure*}

\begin{figure*}
	\centering
	\includegraphics[width=0.9\hsize]{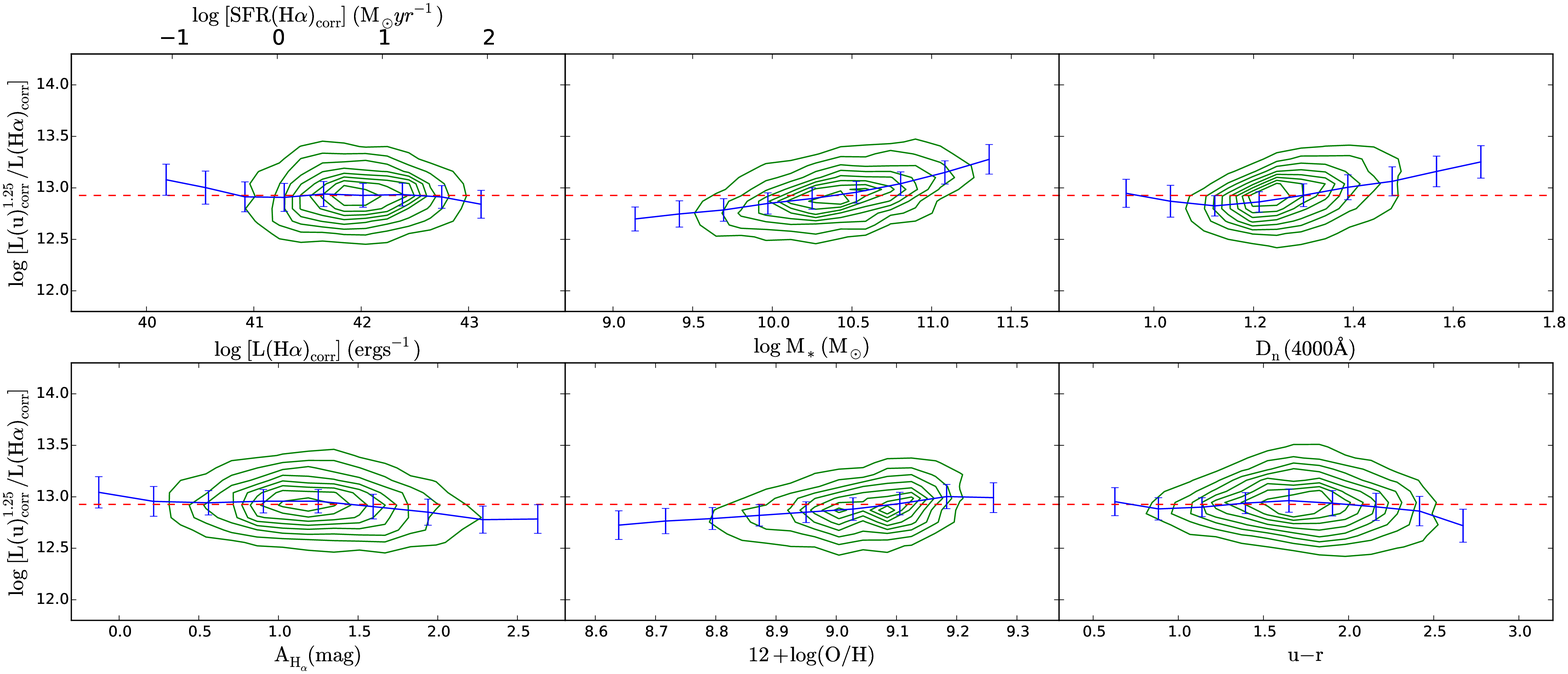}
	\caption{Similar to Figure~\ref{comp1_u_Ha}, but for $L({\it u})_{corr}^{1.25}$ (derived from the nonlinear fitting of Eq.~\ref{eq_noliner_u_ha}) relative to $L(H\alpha)_{corr}$.}
	\label{comp2_u_Ha}
\end{figure*}

The WISE 12 and 22 $\mu$m band luminosities are useful IR tracers of SFRs in galaxies \citep{Shi2012, Lee2013, Wen2014}. We compare the extinction-corrected {\it u}-band luminosities with 12 and 22 $\mu$m band luminosities in Figure \ref{u_vs_wise}. Sample 1 is used to derive the correlation of \itu and 12 $\mu$m, whose best linear and nonlinear fits are
\begin{equation}
	log~L({\it u})_{corr}= log~L(12 \mu m) + (0.50\pm0.13),
\end{equation}
\begin{equation}
	log~L({\it u})_{corr}= (0.74\pm0.01) \times log~L(12 \mu m) + (11.79\pm0.17).
\end{equation}
Sample 2 is used to derive the correlation of \itu and 22 $\mu$m, whose best linear and nonlinear fits are
\begin{equation}
	log~L({\it u})_{corr}= log~L(22 \mu m) + (0.35\pm0.29),
\end{equation}
\begin{equation}
	log~L({\it u})_{corr}= (0.56\pm0.02) \times log~L(22 \mu m) + (19.35\pm0.68).
\end{equation}
The scatters of the nonlinear fits are 0.18 for \itu versus 12 $\mu$m, and 0.27 for \itu versus 22 $\mu$m, larger than that for \itu versus H$\alpha$. All nonlinear and linear coefficients of the relationships between {\it u}-band and \halpha and IR are summarized in Table \ref{tab1}.

\subsection{The Composite SFR Indicators of \itu Band and IR}

Since the starlight absorbed at short wavelengths by interstellar dust is re-radiated in IR, it is possible to use the IR luminosity to trace dust-obscured star formation. Thus, we can use the combination of the IR 12/22 $\mu$m luminosity and observed \itu luminosity to probe the dust-free {\it u}-band luminosity.

Figure \ref{uext_vs_wise} shows the correlation between the attenuation of {\it u}-band (Eq. \ref{eq1}) and WISE 12/22 $\mu$m luminosities, and the best linear fits are plotted:
\begin{equation}
	log~L({\it u})_{att}= log~L(12 \mu m) + (0.26\pm0.14),
    \label{eq_uatt_12}
\end{equation}
\begin{equation}
	log~L({\it u})_{att}= log~L(22 \mu m) + (0.16\pm0.25).
	\label{eq_uatt_22}
\end{equation}
Based on these relations, we can derive the IR-corrected {\it u}-band luminosity:
\begin{equation}
	L({\it u})_{corr,12\mu m}=L({\it u})_{obs}+1.83\times L(12 \mu m)
	\label{IR12u}
\end{equation}
\begin{equation}
	L({\it u})_{corr,22\mu m}=L({\it u})_{obs}+1.46\times L(22 \mu m)
	\label{IR22u}
\end{equation}

Similar to the analyses we did for attenuation-corrected {\it u}-band luminosities in Section \ref{subsec:u_ha}, we compare the IR-corrected {\it u}-band luminosities with the Balmer decrement-corrected \halpha luminosities in Figure \ref{uIR_vs_Ha}. The best linear and nonlinear fits are shown as (also listed in Table \ref{tab1})
\begin{equation}
	log[L({\it u})_{obs}+1.83\times L(12 \mu m)] = log~L(H\alpha)_{corr} + (1.81\pm0.13)
\end{equation}
\begin{equation}
	\begin{split}
		log[L({\it u})_{obs}+1.83\times L(12 \mu m)] = (9.78\pm0.16)\\
		+ (0.81\pm0.01) \times log~L(H\alpha)_{corr} 
		\end{split}
		\label{eq_noliner_uw3_ha}
\end{equation}
\begin{equation}
	log[L({\it u})_{obs}+1.46\times L(22 \mu m)] = log~L(H\alpha)_{corr} + (1.68\pm0.16)
\end{equation}
\begin{equation}
	\begin{split}
		log[L({\it u})_{obs}+1.46\times L(22 \mu m)] = (9.26\pm0.59)\\
		+ (0.82\pm0.01) \times log~L(H\alpha)_{corr}
	\end{split}
	\label{eq_noliner_uw4_ha}
\end{equation}

\section{Discussion}
\label{sec:discussion}

%

\begin{figure*}
	\centering
	\includegraphics[width=0.9\hsize]{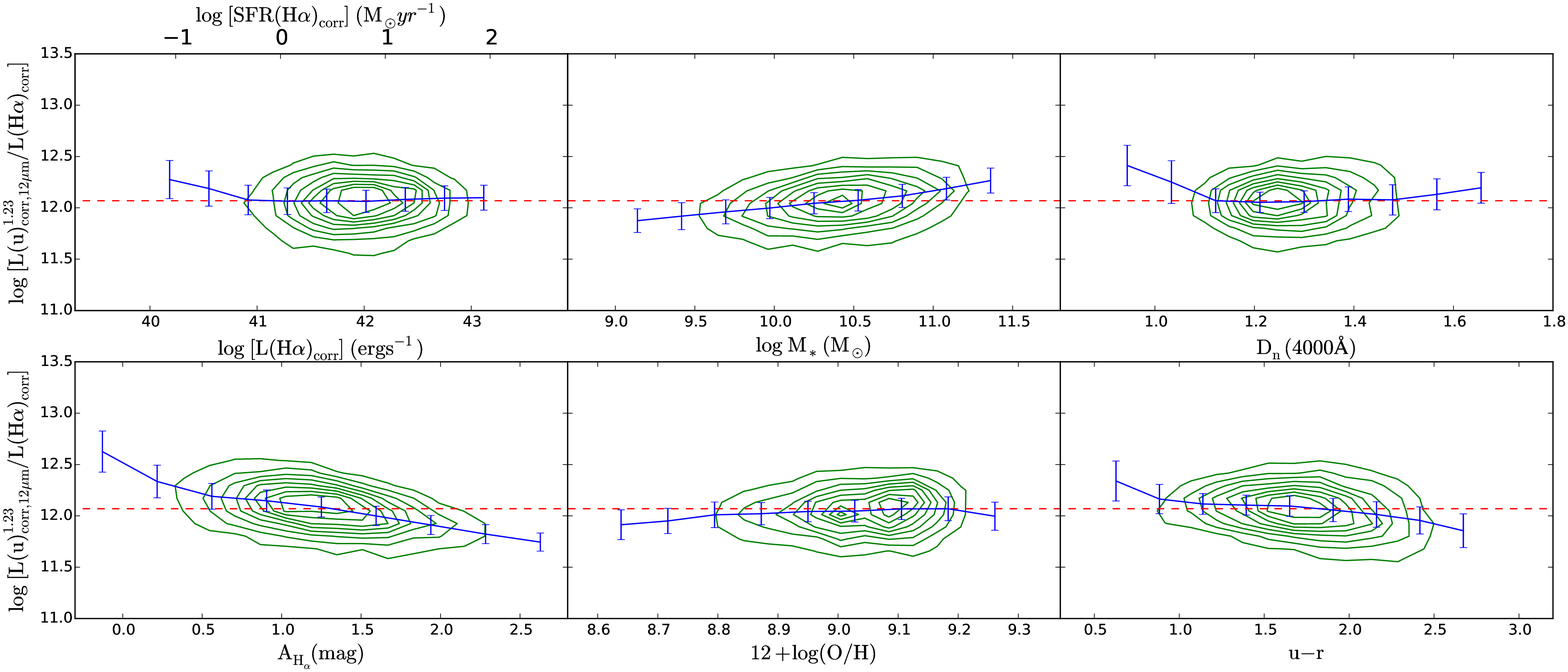}
	\caption{Similar to Figure~\ref{comp1_u_Ha}, but for $L({\it u})_{corr,12\mu m}^{1.23}$ (derived from the nonlinear fit of Eq.~\ref{eq_noliner_uw3_ha}) relative to $L(H\alpha)_{corr}$.}
	\label{comp2_uw3_Ha}
\end{figure*}


\begin{figure*}
	\centering
	\includegraphics[width=0.9\hsize]{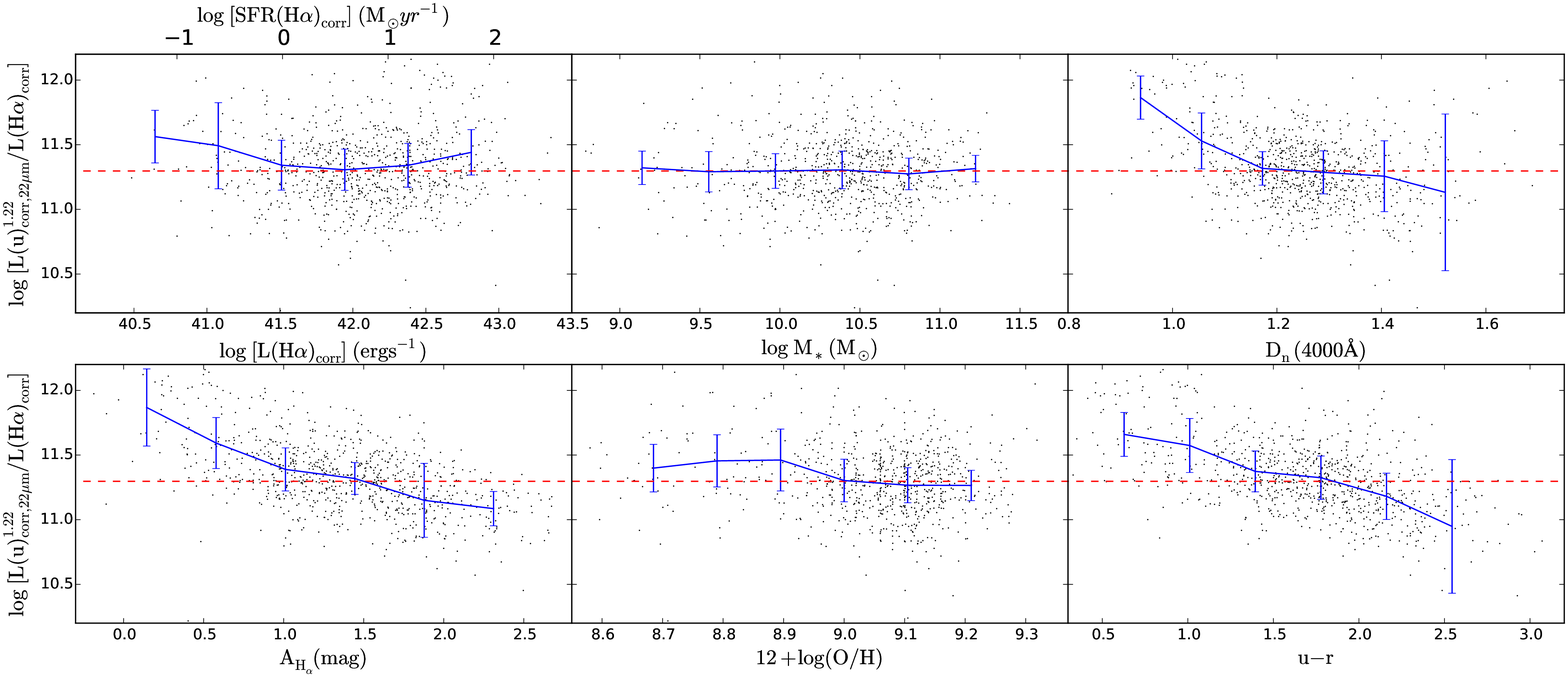}
	\caption{Similar to Figure~\ref{comp1_u_Ha}, but for $L({\it u})_{corr,22\mu m}^{1.22}$ (derived from the nonlinear fit of Eq.~\ref{eq_noliner_uw4_ha}) relative to $L(H\alpha)_{corr}$.}
	\label{comp2_uw4_Ha}
\end{figure*}

In previous sections, we have derived the \itu versus H$\alpha$ relations, and combined \itu and WISE IR luminosities to construct the composite dust-corrected SFR indicators. Specifically, \itu and composite indicators were compared to the aperture- and dust-corrected H$\alpha$ luminosity and were calibrated with linear and nonlinear fits. We have demonstrated that the fits of \itu versus H$\alpha$ with the scatter of $\sim$ 0.17 dex are consistent with the relation of \citet{Hopkins2003}. For the composite indicators, The combinations adopt the form of $L({\it u})_{corr}=L({\it u})_{obs}+\alpha \times L(12\ or\ 22 \mu m)$, and the calibration scatters are 0.16 and 0.19 dex for \itu+12$\mu m$ and \itu+22$\mu m$, respectively.

As we mentioned above, these indicators have been calibrated based on the samples with limited ranges of galaxy properties, it is important to understand their applicability limits and systematic uncertainties and how the uncertainties are affected by galaxy properties.

In Section \ref{sec:sample}, we have described the physical properties of our samples. The present samples are selected using the BPT diagram, they cover the star-forming galaxies in the local universe with the redshift less than 0.4. The ranges of stellar masses, SFRs, and metallicities spanned by our sample galaxies are $10^9$ -- $10^{12} M_{\odot}$, 0.1 -- 100 $M_{\odot} yr^{-1}$, and 12+logO/H = 8.3--9.3, respectively. Therefore, more caution is needed when our calibrations are used to objects outside the above ranges.

In order to establish whether the residuals from these calibrations vary systematically with galaxy properties, we derived systematic residuals against six parameters, i.e., SFRs, stellar masses, $D_n(4000\AA)$, dust attenuation, metallicities, and colors of galaxies. The dependences on these parameters are plotted in Figure~\ref{comp1_u_Ha}--\ref{comp2_uw4_Ha} for the calibrations of each indicators.

In Figure~\ref{comp1_u_Ha} and \ref{comp2_u_Ha}, we present the residuals from the linear and nonlinear calibrations of attenuation-corrected {\it u}-band luminosities (i.e.,  Eq.~\ref{eq_liner_u_ha} and  Eq.~\ref{eq_noliner_u_ha}). Given that H$\alpha$ traces SFRs and $D_n(4000\AA)$ provides a rough measure of the relative age of stellar population in galaxies, we find a dependence of the residuals from the linear fit on SFRs, stellar population, and dust attenuation, such that the residuals decrease with increasing SFRs and attenuations, and increase with the ages of stellar population, while the dependence is very weak for the residuals from the nonlinear fit. These suggest that it is difficult to assign a simple scaling factor to derive an SFR relation for the \itu band \citep{Hopkins2003}, and the nonlinear fits are better than the linear ones. Thus, the nonlinear calibrations are recommended to be used in the measurement of SFRs in galaxies and are mainly focused on in the next analysis. 

The ratios of the {\it u}-band nonlinear fit to H$\alpha$ luminosities in Figure~\ref{comp2_u_Ha} show no or at most marginal trends with nearly all of the physical parameters except weak trends for stellar masses and $D_n(4000\AA)$, which are mainly due to the contamination of old stars to the actual {\it u}-band luminosity \citep{Cram1998}. Galactic color is a rough index of stellar population, which should be used to correct the contribution from old stars in the {\it u}-band \citep[e.g.,][]{K98, Davies2016}. However, we cannot find any clear trend between the ratio and rest-frame color $u-r$ in the right bottom panel of Figure~\ref{comp2_u_Ha}, so it is difficult to find an appropriate form to correct the pollution of old stellar population to the {\it u}-band luminosity using the color $u-r$.

In Figures~\ref{comp2_uw3_Ha} and \ref{comp2_uw4_Ha}, the residuals from the nonlinear fits of combinations \itu+12$\mu m$ and \itu+22$\mu m$ are explored. Similar to the behaviors of the nonlinear fitting result of {\it u}-band luminosity, the ratios of \itu+12$\mu m$ to H$\alpha$ show little trends with the physical parameters at least for most of the samples, while there are weak dependences on stellar masses and dust attenuation. The residuals of \itu+12$\mu m$ show negative correlation with the dust attenuation along with a deviation of 0.1$-$0.3 dex in the range of A$_{H\alpha} <$0.5 mag and A$_{H\alpha} >$2.0 mag. The possible reasons for the correlation are the change in the effective extinction curves, which are affected by star-formation intensity \citep{Hao2011} and the contribution from dust heating by the old stellar population \citep{Bell2003, Zhu2008}. The behaviors of \itu+22$\mu m$ in Figures~\ref{comp2_uw4_Ha} are similar to those of \itu+12$\mu m$ but with larger dispersion, which is probably due to the far fewer galaxies in the sample and the different dust grain species of 12 and 22 $\mu m$ in star-forming galaxies \citep{Zhu2015}.

\section{Summary}
\label{sec:summary}

We have presented and analyzed the possibility of using the optical {\it u}-band luminosities to estimate SFRs of galaxies, and build {\it u}--H$\alpha$ relations. The correlations between {\it u}, H$\alpha$ and IR luminosities were explored based on two samples of BPT-selected star-forming galaxies. The {\it u}-band and the combination of \itu and IR fluxes were empirically calibrated to derive attenuation-corrected SFR indicators. The coefficients in these calibrations are summarized in Table \ref{tab1}, which can be used as the empirical formulae to derive SFRs in normal star-forming galaxies. Our main results and conclusions are summarized as follows.

\begin{enumerate}
\item{} The {\it u}-band luminosities are dust corrected based on Balmer decrements after {\it K}-correction. The attenuation-corrected {\it u}-band luminosities correlate tightly with the Balmer decrement-corrected H$\alpha$ luminosities with the Spearman rank-order correlation coefficient of 0.88. Its nonlinear fit to H$\alpha$ is better than the linear one with an rms scatter of $\sim$ 0.17 dex, consistent with the results of \citet{Hopkins2003}.

\item{} {\it u}-band is compared with WISE 12 and 22 $\mu$m bands, which are useful SFR tracers of galaxies. The rms scatters of their correlations are 0.18 dex for \itu versus 12 $\mu$m, 0.27 dex for \itu versus 22 $\mu$m, larger than that for \itu versus H$\alpha$.

\item{} We compared the correlations between the attenuation of {\it u}-band and WISE 12/22 $\mu$m luminosities, and derived IR-corrected \itu luminosities. Then the linear combinations of 12/22 $\mu$m with {\it u}-band luminosities are calibrated with the Balmer-corrected H$\alpha$ luminosities. The correlation of \itu+12$\mu$m with H$\alpha$ is better than that of \itu+22$\mu$m with H$\alpha$. 

\item{} The systematic residuals of our calibrations are tested against the physical properties of normal star-forming galaxies. The nonlinear fitting results are found to be better than the linear ones, and recommended to be applied in the measurement of SFRs. There are weak dependences on stellar masses, $D_n(4000\AA)$ due to the contamination of old stars, and the residuals of \itu+12/22$\mu$m show slightly negative correlation with the dust attenuation due to the extinction curves and old stellar population.
\end{enumerate}

In brief, this paper provides an attempt of using optical {\it u}-band emission to measure SFRs of galaxies and offers the empirical formulae, while lots of open questions still remain in the calibration of \itu band to galactic SFRs. Therefore, a more detailed analysis of the stellar populations and dust extinction is needed, and the galactic classifications and galaxies at higher redshift will be employed in the future work.

~

\acknowledgements
\label{sec:acknow}
We are grateful to the anonymous referee for thoughtful comments and insightful suggestions that helped to improve this paper.
This project is supported by Chinese National Natural Science Foundation grant Nos. 11303038, 11433005, 11303043, 11373035, and by the National Basic Research Program of China (973 Program), Nos. 2014CB845704, 2014CB845702, and 2013CB834902, and also by the Strategic Priority Research Program ``The Emergence of Cosmological Structures'' of the Chinese Academy of Sciences, grant No. XDB09000000. This project is also supported by the Young Researcher Grant of National Astronomical Observatories, Chinese Academy of Sciences. 

The SCUSS project is funded by the Main Direction Program of Knowledge Innovation of Chinese Academy of Sciences (No. KJCX2-EWT06). It is also an international cooperative project between National Astronomical Observatories, Chinese Academy of Sciences, and Steward Observatory, University of Arizona, USA. Technical support and observational assistance from the Bok telescope are provided by the Steward Observatory. The project is managed by the National Astronomical Observatory of China and Shanghai Astronomical Observatory. The management and publication of data are supported by the Chinese Astronomical Data Center and the China-VO team.

This research makes use of the cross-match service provided by CDS, Strasbourg. It makes use of data products from the {\it Wide-field Infrared Survey Explorer}, which is a joint project of the University of California, Los Angeles, and the Jet Propulsion Laboratory/California Institute of Technology, funded by the National Aeronautics and Space Administration. Funding for the SDSS and SDSS-II has been provided by the Alfred P. Sloan Foundation, the Participating Institutions, the National Science Foundation, the U.S. Department of Energy, the National Aeronautics and Space Administration, the Japanese Monbukagakusho, the Max Planck Society, and the Higher Education Funding Council for England. The SDSS website is http://www.sdss.org/. The SDSS MPA-JHU catalog was produced by a collaboration of researchers (currently or formerly) from the MPA and the JHU. The team is made up of Stephane Charlot, Guinevere Kauffmann and Simon White (MPA), Tim Heckman (JHU), Christy Tremonti (University of Arizona - formerly JHU), and Jarle Brinchmann (Centro de Astrof\'isica da Universidade do Porto - formerly MPA).

~

~




\end{CJK}
\end{document}